\documentclass{ws-procs9x6}

\begin{document}

\title{Effective $\bar{K}N$ interaction and role of chiral symmetry}

\author{Tetsuo Hyodo$^*$}

\address{Department of Physics, Tokyo Institute of Technology 
 Meguro 152-8551, Japan \\
$^*$E-mail: hyodo@th.phys.titech.ac.jp}

\author{Wolfram Weise}

\address{Physik-Department, Technische Universit\"at M\"unchen, 
D-85747 Garching, Germany}

\begin{abstract}
    We study the consequence of chiral SU(3) symmetry in the kaon-nucleon 
    phenomenology, by deriving the effective single-channel $\bar{K}N$ potential.
    It turns out that the $\pi\Sigma$ interaction is strongly attractive and 
    plays an important role for the structure of the $\Lambda(1405)$ resonance.
    We discuss the implication of effective potential for the few-body kaonic 
    nuclei.
\end{abstract}

\keywords{Chiral SU(3) dynamics; $\Lambda(1405)$ resonance; 
Kaonic nuclei.}

\bodymatter

\section{Introduction}\label{sec:intro}

Physics of $\bar{K}$ in nuclei is now lively discussed, stimulated by the 
proposal of deeply bound kaonic nuclei~\cite{Akaishi:2002bg}. Experimental 
searches for the simplest $K^-pp$ system have reported a broad bump structure in 
the $\Lambda N$ invariant mass spectrum, while the interpretation of the observed
structure requires an elaborate analysis~\cite{Kppexp}. Rigorous few-body 
calculations with realistic potentials for the $K^-pp$ system were presented in 
Refs.~\refcite{Kppth}, indicating the bound $K^-pp$ system above $\pi\Sigma N$ 
threshold.

The relevant energy region for the study of kaonic nuclei is far below the 
$\bar{K}N$ threshold, where the amplitude is dominated by the $\Lambda(1405)$ 
resonance. This means that the extrapolation of the $\bar{K}N$ interaction should
be carefully performed with a proper treatment of the $\Lambda(1405)$. In this 
study, we rely upon the following guiding principles for the description of the 
$\bar{K}N$ scattering: chiral symmetry and coupled-channel dynamics. Chiral 
symmetry of QCD determines the low energy interaction between the pseudoscalar 
meson (the Nambu-Goldstone boson) and any target hadron~\cite{WT}, 
and the importance of the coupled-channel dynamics has been emphasized in the 
phenomenological study of $\bar{K}N$ scattering, based on the strong attraction 
in this channel~\cite{Dalitz:1967fp}. Theoretical framework based on these 
principles has been developed as the chiral coupled-channel approach, reproducing
successfully the $\bar{K}N$ scattering data and the properties of the 
$\Lambda(1405)$ resonance~\cite{ChU}. Here we derive an effective $\bar{K}N$ 
interaction based on chiral dynamics~\cite{Hyodo:2007jq}, and discuss its 
phenomenological consequence in the study of $K^-pp$ system~\cite{DoteKpp}.

\section{Chiral interaction and coupled-channel approach}\label{sec:interaction}

In the leading order of chiral perturbation theory, meson-baryon $s$-wave 
interaction at total energy $\sqrt{s}$ from channel $j$ to $i$ reads
\begin{align}
    V_{ij}(\sqrt{s})
    =&-\frac{C_{ij}}{4f^{2}}
    (2\sqrt{s}-M_i-M_j)\sqrt{\frac{E_i+M_i}{2M_i}}
    \sqrt{\frac{E_j+M_j}{2M_j}} 
    \label{eq:WTint} ,
\end{align}
where $f$ is the pseudoscalar meson decay constant, $M_i$ and $E_i$ are the mass 
and the energy of the baryon in channel $i$, respectively. The coupling strengths
$C_{ij}$ are collected in the matrix
\begin{equation}
    C_{ij}^{I=0}
    =\begin{pmatrix}
       3 & -\sqrt{\frac{3}{2}} & \frac{3}{\sqrt{2}} & 0 \\
         & 4 & 0 & \sqrt{\frac{3}{2}} \\
	 &   & 0 & -\frac{3}{\sqrt{2}} \\
	 &   &   & 3
    \end{pmatrix}
    \nonumber ,
\end{equation}
for the $S=-1$ and $I=0$ channels in the following order : $\bar{K}N$, 
$\pi\Sigma$, $\eta\Lambda$, and $K\Xi$. The properties of the interaction---sign,
strength, and energy dependence---are strictly governed by the chiral low energy 
theorem. One observes that the interactions in \textit{both} $\bar{K}N$ and 
$\pi\Sigma$ channels are attractive, which is inevitable as far as we respect 
chiral symmetry. As we will see below, these attractive forces are so strong that
pole singularities of the amplitude are generated for both channels. 

Since the system is strongly interacting, we need to perform nonperturbative
resummation. In Refs.~\refcite{ChU,Hyodo:2007jq} this has been achieved by 
solving the Bethe-Salpeter equation
\begin{align}
    T_{ij}(\sqrt{s})
    =& 
    V_{ij}(\sqrt{s})+V_{il}(\sqrt{s})\,G_{l}(\sqrt{s})\,T_{lj}(\sqrt{s}),
    \label{eq:full}
\end{align}
with the interaction kernel $V_{ij}$ in Eq.~\eqref{eq:WTint} and the 
meson-baryon loop integral $G_i$ in dimensional regularization. The solution 
of Eq.~\eqref{eq:full} is given in matrix form by $T= [V^{-1}-G]^{-1}$ under 
the on-shell factorization. The equivalent amplitude is also obtained in the N/D 
method, which guarantees the unitarity of the scattering amplitude. 

\section{Structure of the $\Lambda(1405)$ resonance}\label{sec:L1405}

It has been shown that the chiral coupled-channel approach reproduces the 
experimental observables of $\bar{K}N$ scattering very well, generating the 
$\Lambda(1405)$ resonance dynamically~\cite{ChU,HNJH}. The $\Lambda(1405)$ in 
this approach is naively interpreted as quasibound meson-baryon molecule. Indeed,
several recent analyses support the meson-baryon molecular picture\cite{L1405}.

In the present context, it is important to focus on the pole structure. In 
Ref.~\refcite{Jido:2003cb}, it is found that the $\Lambda(1405)$ resonance is 
associated by two poles. Using the model given in Refs.~\refcite{HNJH}, the poles
are found at
\begin{align*}
    z_1 &= 1428 - 17 i \text{ MeV}, \quad
    z_2 = 1400 - 76 i \text{ MeV} ,
\end{align*}
which appear above $\pi\Sigma$ threshold and below $\bar{K}N$ threshold. 
Since the two poles are located close to each other, the observed spectrum 
exhibits only one bump structure, which was interpreted as a single 
resonance, the $\Lambda(1405)$. The coupling strengths of the poles to the 
$\pi\Sigma$ and $\bar{K}N$ channels are different from one to the other. 
Therefore, these poles contribute to the $\bar{K}N$ and $\pi\Sigma$ 
amplitudes with different weights, leading to the different spectral shapes 
of two amplitudes~\cite{Jido:2003cb}. 

\begin{figure}[tbp]
    \centering
    \includegraphics[width=0.5\textwidth,clip]{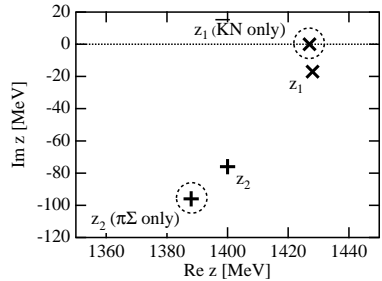}
    \caption{\label{fig:pole}
    Pole positions of the results of single-channel resummation 
    [$z_1(\bar{K}N$ only) and $z_2(\pi\Sigma$ only)] together with the poles
    in the $\bar{K}N(I=0)$ full coupled-channel amplitude ($z_1$ and $z_2$).}
\end{figure}

In order to study the origin of this interesting structure, 
we perform the resummation of the single-channel interaction by switching off
the couplings to the other channels. This single channel $\bar{K}N$ 
interaction generates a relatively weak bound state below threshold, while 
the $\pi\Sigma$ amplitude exhibits a broad resonance above threshold:
\begin{equation}
    z_1(\bar{K}N \text{ only}) = 1427  \text{ MeV} ,
    \quad z_2(\pi\Sigma \text{ only}) = 1388 - 96 i \text{ MeV} .
    \nonumber
\end{equation}
Thus, the attractive forces in diagonal $\bar{K}N$ and $\pi\Sigma$ channels 
already generate two poles between thresholds. We plot the positions of these 
poles in Fig.~\ref{fig:pole}, together with the poles of the full amplitude in 
the coupled-channel framework. The figure obviously suggests that the pole 
$z_1(\bar{K}N \text{ only})$ is the origin of the pole $z_1$, whereas 
$z_2(\pi\Sigma \text{ only})$ evolves to the pole $z_2$. This observation agrees 
with the qualitative behavior discussed in Ref.~\refcite{Jido:2003cb}; the pole 
$z_1$ strongly couples to the $\bar{K}N$ channel and the pole $z_2$ to the 
$\pi\Sigma$ channel.

It is interesting to note that the higher energy $\bar{K}N$ channel has stronger 
attraction to generate a bound state, and the lower energy $\pi\Sigma$ channel 
shows the relatively weaker attraction, which is nevertheless strong enough to 
create a resonance. The appearance of the two poles in this energy region is 
caused by the balance of the two attractive forces. Similar pole structure is 
observed in the meson-meson scattering sector: $\sigma$ and $f_0(980)$ resonances
in $\pi\pi$-$\bar{K}K$ system (flavor structure is the same). In this sense, the 
physics of the lower energy pole is related to the $\sigma$ meson through chiral 
symmetry.

\section{Effective single-channel potential and $K^-pp$ system}\label{sec:single}

Keeping the structure of the $\Lambda(1405)$ in mind, we construct an effective 
single-channel $\bar{K}N$ interaction which incorporates the dynamics of the 
other channels 2-4 ($\pi\Sigma$, $\eta N$, and $K\Xi$). We would like to obtain 
the solution $T_{11}$ of Eq.~\eqref{eq:full} by solving a single-channel equation
with kernel interaction $V^{\text{eff}}$,
\begin{align}
    T^{\text{eff}}
    =& \,V^{\text{eff}}+V^{\text{eff}}\,G_1\,T^{\text{eff}}
    = \,T_{11} .
    \nonumber 
\end{align}
Consistency with Eq.~\eqref{eq:full} requires that $V^{\text{eff}}$ be the sum of
the bare interaction in channel 1 and the contribution from other channels (2-4),
which includes iterations of one-loop terms to all orders. This method is exactly
equivalent to the original coupled-channel framework, as far as the $\bar{K}N$ 
scattering amplitude is concerned.

In this way, the effective $\bar{K}N$ interaction $V^{\text{eff}}$ is calculated 
with a chiral coupled-channel model~\cite{HNJH}. It turns out that the 
$\pi\Sigma$ and other coupled channels enhance the strength of the interaction at
low energy, although not by a large amount. The primary effect of the coupled 
channels is found in the energy dependence of the interaction kernel. By 
calculating the scattering amplitude, it is found that the resonance structure in
the $\bar{K}N$ channel is observed at around 1420 MeV, higher than the nominal 
position of the $\Lambda(1405)$. The deviation of the resonance position has a 
large impact to the single-channel $\bar{K}N$ potential, which should be 
constructed so as to reproduce the scattering amplitude in $\bar{K}N$ channel. 
Measuring from the $\bar{K}N$ threshold ($\sim 1435$ MeV) we find the ``binding 
energy'' of the $\Lambda(1405)$ as $\sim 15$ MeV, not as the nominal value of 
$\sim 30$ MeV. This shift of the binding energy apparently reduces the strength 
of the potential.

Next we construct an equivalent local $\bar{K}N$ potential in coordinate space. 
We consider an $s$-wave antikaon-nucleon system in nonrelativistic quantum 
mechanics through the Schr\"odinger equation with a potential $U(r,E)$. As 
explained in detail in Ref.~\refcite{Hyodo:2007jq}, the local potential $U(r,E)$ 
has been constructed such that the scattering amplitude in coupled-channel 
approach is reproduced in this system. This is not an exact transformation, since
it is not guaranteed that a simple local potential can reproduce the complicated 
coupled-channel dynamics. Nevertheless, we have constructed a complex and 
energy-dependent $\bar{K}N$ potential with the gaussian form of the spatial 
distribution, which well reproduces the coupled-channel results.

When the potential is applied to the variational calculation of $K^-pp$ system 
together with the realistic NN interaction~\cite{DoteKpp}, bound state solution 
is found with a smaller binding energy than the other calculations~\cite{Kppth}.
The main reason for the small binding is the weaker attraction of the effective 
$\bar{K}N$ potential. In our approach, chiral symmetry requires the strong 
$\pi\Sigma$ dynamics, and the attraction force to form the $\Lambda(1405)$ is 
divided into $\bar{K}N$ and $\pi\Sigma$ channels. As a consequence, the allotment
of the $\bar{K}N$ attraction is rather small. 

Present analysis focuses on the $\bar{K}NN$ component, since the $\pi\Sigma N$ 
channel is eliminated from the model space. For much lower energy region close to
the $\pi\Sigma N$ threshold, explicit treatment of the $\pi\Sigma N$ channel 
would play an important role\cite{Ikeda:2008ub}, which may be related to the 
lower energy pole in the $\bar{K}N$-$\pi\Sigma$ amplitude. In addition, explicit 
treatment of $\Lambda N$ channel is mandatory, in order to compare the 
theoretical prediction of the bound state with the experimentally observed bump 
structure in the $\Lambda N$ spectrum~\cite{Kppexp}

\section{Summary and perspective}\label{sec:summary}

We have derived an effective $\bar{K}N$ interaction based on chiral low 
energy theorem and the coupled-channel dynamics. We show that the chiral 
interaction leads to a strongly interacting $\pi\Sigma$-$\bar{K}N$ system, in
which the $\Lambda(1405)$ is described as the $\bar{K}N$ quasibound state 
embedded in the \textit{resonating} $\pi\Sigma$ continuum. We construct an 
equivalent local potential in single $\bar{K}N$ channel, which represents the
effect of coupled-channel dynamics through the imaginary part and energy 
dependence. As a consequence of the strong $\pi\Sigma$ dynamics, the 
resulting potential is less attractive than the purely phenomenological 
potential in the subthreshold energy range.

It is worth emphasizing that there is no direct experimental constraint on 
the $\bar{K}N$ amplitude below threshold. We have to extrapolate $\bar{K}N$ 
interaction calibrated by scattering data above threshold, down to the 
relevant energy scale. Here we utilize the principle of chiral SU(3) symmetry
in order to reduce the ambiguity of the extrapolation. Comprehensive study of the
threshold $\bar{K}N$ data and the spectrum of the $\Lambda(1405)$ should play an 
important role to constraint the $\bar{K}N$ interaction below threshold.

\section*{Acknowledgments}

This project is partially supported by BMBF, GSI, by the DFG excellence cluster 
``Origin and Structure of the Universe.", by the Japan Society for the Promotion 
of Science (JSPS), and by the Grant for Scientific Research (No.\ 19853500) from 
the Ministry of Education, Culture, Sports, Science and Technology (MEXT) of 
Japan. T.H. thanks the support from the Global Center of Excellence Program
by MEXT, Japan through the Nanoscience and Quantum Physics Project of the Tokyo 
Institute of Technology.

\end{document}